\documentclass[aps,preprintnumbers,amsmath,amssymb,prl,superscriptaddress,longbibliography,twocolumn]{revtex4-2}
\usepackage{dcolumn}
\usepackage{color}
\usepackage[caption=false]{subfig}
\usepackage{feynmp}
\usepackage{feynmp-auto}
\usepackage{systeme,mathtools}
\usepackage{stmaryrd}
\usepackage{easyReview}
\usepackage[normalem]{ulem}

\usepackage{mathtools}
\usepackage{nicefrac}
\usepackage{soul}

\usepackage{tikz}
\usetikzlibrary{decorations.pathmorphing,arrows.meta}

\def\comment#1{}

\newcommand{\nc}{\newcommand}
\nc{\scs}{\scriptstyle}
\nc{\setval}{\fmfset{wiggly_len}{3mm} \fmfset{arrow_len}{1.5mm}
	\fmfset{arrow_ang}{13} \fmfset{dash_len}{1.5mm}\fmfpen{0.125mm}
	\fmfset{dot_size}{2thick}}

\usepackage{bm,latexsym,mathrsfs,enumerate,color}
\usepackage[mathcal]{euscript}
\usepackage[breaklinks=true,unicode=true,urlcolor = purple,colorlinks = true,citecolor = purple,linkcolor = purple]{hyperref}
\usepackage{graphicx}
\usepackage{todonotes}
\usepackage{wrapfig}
\usepackage{easyReview}

\renewcommand{\vec}[1]{\bm{#1}}

\def\slashchar#1{\setbox0=\hbox{$#1$}           
	\dimen0=\wd0                                 
	\setbox1=\hbox{/} \dimen1=\wd1               
	\ifdim\dimen0>\dimen1                        
	\rlap{\hbox to \dimen0{\hfil/\hfil}}      
	#1                                        
	\else                                        
	\rlap{\hbox to \dimen1{\hfil$#1$\hfil}}   
	/                                         
	\fi}                                         %

\DeclareMathAlphabet\mathbfcal{OMS}{cmsy}{b}{n}

\def\nablab{{\mbox{\boldmath $\nabla$}}}

\def\gammab{{\mbox{\boldmath $\gamma$}}}

\makeatletter
\newcommand{\customlabel}[2]{%
\protected@write \@auxout {}{\string \newlabel {#1}{{#2}{}}}}

\makeatother

\begin{document}
	
	\title{Unconventional criticality in $O(D)$-invariant loop-constrained Landau theory}
	
	\author{Svitlana Kondovych}
	\affiliation{Institute for Theoretical Solid State Physics, Leibniz Institute for Solid State and Materials Research Dresden, Helmholzstr. 20, D-01069 Dresden, Germany}
	
	\author{Asle Sudb\o}
	\affiliation{Center for Quantum Spintronics, Department of Physics, Norwegian University of Science and Technology, NO-7491 Trondheim, Norway}
	
	\author{Flavio S. Nogueira}
	\affiliation{Institute for Theoretical Solid State Physics, Leibniz Institute for Solid State and Materials Research Dresden, Helmholzstr. 20, D-01069 Dresden, Germany}
	
	\begin{abstract}
        We study an unconventional phase transition in ferroelectrics where the polarization field is constrained to be divergence-free, allowing only loop-like configurations. This local constraint fundamentally alters the critical behavior, driving the system beyond the Landau-Ginzburg-Wilson paradigm. A renormalization group analysis shows that the polarization acquires an unusually large anomalous dimension, $\eta\approx 0.239$ in three dimensions, far exceeding the typical values in $O(3)$-invariant systems. We attribute this effect to a naturally induced gauge symmetry originating from the zero divergence constraint. Such constraint-induced gauge-field behavior is reminiscent of fractionalized phases, revealing a fundamental connection between constrained ferroelectrics and emergent gauge phenomena in correlated matter.
	\end{abstract}

	\maketitle

\textit{Introduction --} The Landau-Ginzburg-Wilson (LGW) paradigm of phase transitions, built upon the concept of spontaneous symmetry breaking and local order parameters, has been a cornerstone of condensed matter physics. However, numerous systems have revealed the limitations of this framework, triggering the exploration of critical phenomena that go beyond conventional symmetry-breaking descriptions. These include transitions driven by topological order, quantum entanglement, and fractionalized excitations, as seen in quantum spin liquids and deconfined quantum criticality (DQC)~\cite{Senthil2004,Sachdev2011,Prakash2025,Senthil_PhysRevB.70.144407}. Such transitions challenge our understanding of universality and scaling, offering insights into complex phases of matter with potential applications in quantum computation, quantum information, and quantum sensing, as well as in materials science.

In particular, systems subjected to local constraints offer a promising platform for unconventional criticality. A prominent example involves {\it divergence-free} conditions on vector fields, which manifest across various physical systems. For instance, in
frustrated antiferromagnets on pyrochlore and kagome lattices~\cite{Henley2010}, divergence-free constraints on staggered magnetization or local flux variables emerge from local constraints imposed by competing interactions. Such a constraint also arises in theories of three-dimensional quantum dimer models~\cite{Moessner2001,Nogueira2009}.  

In continuum systems, magnetostatics offers a classical example, where the solenoidal constraint on magnetic field, $\nablab\cdot\vec{B}=0$, governs flux conservation and enforces loop-like field configurations~\cite{Jackson1998}. This imposes topological restrictions on magnetic fluids, leading to conserved magnetic helicity and the formation of coherent, filamentary magnetic structures~\cite{Freidberg2014}.
In the context of incompressible fluids, the velocity field $\vec{v}$ satisfies $\nablab\cdot\vec{v}=0$, leading to nonlocal interactions through pressure mediation. According to  Arnold's theorem for incompressible liquids~\cite{Arnold2021}, the stationary fluid flow in this case is composed of two types of elementary topological excitations: either vortex-like or knotted hopfion-like ones. Similarly, in nanoscale ferroelectrics, e.g. in SrTiO$_3$/PbTiO$_3$ superlattices or PbTiO$_3$ and PbZr$_{0.6}$Ti$_{0.4}$O$_3$ nanoparticles~\cite{Lukyanchuk2020,Kondovych2023,DiRino2025,Lukyanchuk2025}, a zero divergence constraint on the polarization field $\nablab\cdot\vec{P}=0$ plays a central role in the formation of vortex- and hopfion-like textures, in which the polarization field streamlines wind around nested cylinders or tori. This prevents the formation of polarization bound charges, $\rho_p=-\nablab\cdot\vec{P}$, which arise due to the nonuniform distribution of polarization dipoles, and minimizes the huge electrostatic energy associated with depolarization effects. 
Such constraint-driven structuring of field lines gives rise to rich topological phenomena~\cite{Guo2022,Wang2023,Lukyanchuk2024,Lukyanchuk2025}, with analogies to hydrodynamics and emergent gauge theories. Similarly, in a superfluid, the constraint $\nablab\cdot\vec{v}_s=0$ on the superfluid velocity $\vec{v}_s$ ultimately leads to vortex loop excitations~\cite{Feynman1955,Fetter1967,Onsager1949}, which determine its phase structure~\cite{Shenoy1989,Nguyen1998}. In particular, in two dimensions, where vortices are points, this point of view leads to the celebrated Berezinskii-Kosterlitz-Thouless (BKT) phase transition~\cite{Berezinskii1971,Kosterlitz1973}. 

These are just a few examples of the diverse classical and quantum systems where divergence-free conditions govern the low-energy physics, generating gauge structures and long-range correlations. When supplemented with continuous symmetries such as the orthogonal group $O(D)$, where $D$ is the spatial dimension, these systems can exhibit LGW-incompatible critical behavior, which is the focus of this work. Crucially, we address the role of the functional measure of the partition function as a significant factor altering the universality class of the system. This means that the phase structure of the system is not only determined by the action and its symmetries. As an exemplary system revealing such a behavior, we analyze ferroelectric materials where the polarization is constrained to be divergence-free, $\nablab\cdot\vec{P}=0$, giving rise to topological configurations that influence the system's critical properties.
We refer to such materials as topological ferroelectrics. This constraint is precisely accounted for in the integration measure of the partition function.

A natural question emerging from the above considerations concerns the role played by the zero divergence constraint in determining the phase structure of the physical system. How much does such a constraint affect the LGW theory of a ferroelectric system? Remarkably, we find that the critical behavior of correlation functions is affected significantly, reflecting a scenario that is usually encountered in physical systems exhibiting fractionalization. The latter is a property underlying the decay of an order parameter into emergent building blocks, similarly to the physics of quark deconfinement as studied in high-energy physics. In condensed matter physics, this is generally associated to an emergent gauge symmetry, with realizations in quantum antiferromagnets featuring competing orders. This leads to a phase structure incompatible with the LGW paradigm~\cite{Senthil2004,Senthil_PhysRevB.70.144407}.  
One of the most salient features underlying this scenario is that it leads to a unusually large anomalous dimension, $\eta$, of the order parameter $O(\vec{x})$~\cite{Senthil2004}. Recall that LGW-compatible systems have considerably smaller values of $\eta$ compared to those of thermodynamic critical exponents, namely, exponents obtained from the scaling of the free energy density~\cite{Kleinert2001,zinn2021quantum}.

The route towards non-LGW behavior in this work differs substantially from the DQC mechanism, as it does not involve fractionalization and keeps the LGW action intact. However, in spite of this, the partition function is changed in an essential way due to a constrained integral measure on the fluctuations. This in turn considerably affects the critical behavior of the order parameter correlation functions, leading to an unusually large anomalous dimension, similar to DQC. Specifically, we show that accounting for the constraint $\nablab\cdot\vec{P}=0$ for the electric polarization indeed implies a large anomalous dimension without apparent fractionalized excitations. In fact, we find at two-loop order in perturbation theory that $\eta=0.239$ in three dimensions. By contrast, in a LGW theory with $O(N)$ symmetry one finds a value of $\eta$ which is typically one order of magnitude smaller than the value found here. Indeed, in an $O(3)$ symmetric theory, which is the natural case to compare with in the present context, the result is $\eta\approx 0.034$. Crucially, the constraint $\nablab\cdot\vec{P}=0$ naturally imposes that $N=D$, i.e., the number of components of the order parameter coincides with spatial dimensionality.

\textit{Constraint-generated loops --}
The zero divergence constraint in topological ferroelectrics can be viewed from a statistical field theory perspective as representing the collective behavior of objects that we call here {\it polarization loops}, or {\it$P$-loops}, which have many aspects in common with vortex loops in three-dimensional superfluids. At this point it is worth mentioning the well known dielectric approach to understand phase transitions in two-dimensional superfluids~\cite{Minnhagen1987}. While three-dimensional superfluids feature vortex loop excitations~\cite{Feynman1955,Fetter1967,Onsager1949}, in two dimensions vortices are point-like objects that arise in pairs having opposite vorticities ($+$ or $-$)~\cite{Kosterlitz1973}. This is more easily seen by observing that a vortex loop pierces a plane at two points. The point where the loop flows to corresponds to (say) positive vorticity, the one where it outflows yields a negative vorticity. Hence, an ensemble of point vortices in a two-dimensional superfluid can be viewed as a Coulomb gas~\cite{Frohlich1981} where the vortices represent electric charges. Neutrality of the Coulomb gas is guaranteed by the fact that point vortices arise only in pairs of opposite vorticities~\cite{Minnhagen1987}. A two-dimensional Coulomb gas undergoes a phase transition from a dielectric phase where vortex pairs are bound into dipoles to a plasma phase where the dipoles unbind. This is the physical picture behind the BKT phase transition. In the context of this work, such a Coulomb gas is actually a dual representation of the two-dimensional $O(2)$ nonlinear $\sigma$-model ($N=D=2$). As a model for a two-dimensional superfluid, the latter is also subject to a zero divergence constraint, namely, the superfluid velocity $\vec{v}_s$ satisfies the incompressibility condition $\nablab\cdot\vec{v}_s=0$.  

An attempt to implement the same point of view in the case of a three-dimensional Coulomb gas faces a major challenge. In fact, unlike its two-dimensional counterpart, the latter does not undergo any phase transition, remaining permanently in the plasma phase. This occurs because dipoles in three dimensions do not screen~\cite{Frohlich1981,Kragset2004}. A nontrivial way to have a Coulomb system with spontaneous symmetry breaking is constraining the polarization charge density $\rho_p$ to vanish, meaning $\rho_p=-\nablab\cdot\vec{P}=0$, realized either through the described streamline winding or through surface compensation mechanisms~\cite{Efe2021}.  

To extend this dielectric point of view to higher dimensions, we represent the electric polarization as, 
\begin{equation}
	\label{Eq:P-loop}
	\vec{P} \left( \vec{r} \right) = \sum_a  q_a \int_{L_a}  d \gammab^{(a)} \delta^D \left( \vec{r} - \gammab^{(a)} \right),
\end{equation}
where $q_a$ are electric charges and $L_a$ is a path associated to a given charge. If $\nablab\cdot\vec{P}=0$, all the paths form loops, akin to vortex loops in three-dimensional superfluids. Interestingly, a rotational loop-like dipole arrangement leads to a so called ferroaxial behavior~\cite{Johnson2012,Hlinka2016,Hayashida2020}. The experimental evidence of a ferroaxial density wave has been recently reported in connection with an axial Higgs mode in rare-earth tritellurides ErTe$_3$ and HoTe$_3$~\cite{Burch2025}. In this sense, in addition to the polar ordering under the constraint $\nablab\cdot\vec{P}=0$, a secondary ferroaxial order is present, enforced by the same constraint and characterized by an electric toroidal (ferroaxial) moment $\frac{1}{2}\,\vec{r}\times\vec{P}$.

Using a parameter $s\in[0,1]$ such that $\gammab^{(a)}(s=0)=\gammab^{(a)}(s=1)$, we obtain that the total dipole moment associated with loops,
\begin{equation}
    \vec{p}=\sum_a q_a\int_0^1 ds\frac{d\gammab^{(a)}}{ds}=0,
\end{equation}
vanishes in this representation, while the ferroaxial, angular momentum-like vector, 
\begin{equation}
\label{Eq:FA-L}
    \vec{L}_{\rm FA}=-\frac{1}{2}\sum_a q_a\int_0^1 ds\gammab^{(a)}(s)\times \frac{d\gammab^{(a)}}{ds},
\end{equation}
does not vanish in general. 
We should note that the vanishing of the total dipole moment does not occur when thermal fluctuations are accounted for, as we demonstrate further in this work.

Remarkably, the $P$-loop representation allows for an elegant proof (see Appendix~\hyperlink{app:Helicity}{\ref{app:Helicity}}) of the equivalence between the average linking of the polarization field lines, $\mathscr{L}_{\vec{P}}$, and the helicity (of Hopf invariant) of the field, $\mathscr{H}$, known as Arnold's helicity theorem~\cite{Arnold2021} in application to ferroelectrics~\cite{Lukyanchuk2025}. It highlights the topological nature of the polarization under the zero divergence constraint.

To study the critical properties of the presented $P$-loops, we employ an LGW type of theory where the constraint $\nablab\cdot\vec{P}=0$ is enforced, in a situation characteristic of topological ferroelectrics. Importantly, although our model is LGW-based, the divergence-free constraint removes the standard assumption that the order parameter components fluctuate independently and subjected only to the internal symmetries of the action. In the case considered here the constraint must be incorporated in the integral measure of the partition function. This places the description beyond the LGW paradigm, because in the study of critical fluctuations it is the partition function rather than the action that matters. The latter still has a LGW form though, but integrating over the fields in the partition function requires changing the integral measure to accommodate the local constraint $\nablab\cdot\vec{P}=0$.

\textit{Renormalization group analysis -- }In  the absence of anisotropies, the action of a topological ferroelectric is given in $D$ dimensions by~\cite{Landau2013}, 
\begin{equation}
	\label{Eq:S-Ferroelectric}
	S=\int d^Dr\left[\frac{1}{2}(\nablab\vec{P})^2+\frac{a}{2}\vec{P}^2+\frac{b}{8}(\vec{P}^2)^2\right],
\end{equation}
where $\vec{P}$ is a $D$-dimensional polarization vector, see Eq. (\ref{Eq:P-loop}), and the constraint $\nablab\cdot\vec{P}=0$ holds, arising from the absence of depolarization charges in the low-energy states. This implies that only closed-loop configurations of polarization corresponding to bound dipole textures are energetically favored and contribute to the partition function. Recall that in dielectrics the polarization couples linearly to the electric field, which implies that the ``mass'' coefficient $a$ appearing in the action corresponds to the inverse susceptibility. Ferroelectric order, on the other hand, causes the electric field to depend nonlinearly in the polarization, so higher powers of $\vec{P}$ arise in the expression of the electric field, and ferroelectricity is induced by the  sign change in $a$ as the temperature is reduced below the transition temperature, thus driving the onset of a spontaneous polarization. The coefficient $b>0$ physically controls the stability of the energy in the spontaneous symmetry breaking process~\cite{Landau2013}.

In order to study the critical fluctuations of the system~(\ref{Eq:S-Ferroelectric}), we consider the partition function, 
\begin{equation}
	\label{Eq:Z}
\mathcal{Z}=\int\mathcal{D}\vec{P}\delta(\nablab\cdot\vec{P})e^{-S},
\end{equation} 
where the constraint is incorporated into the path integral measure. Using a functional integral representation of the delta function, a modified action is obtained featuring a Lagrange multiplier field $\lambda$ enforcing the constraint, 
\begin{equation}
	\label{Eq:S-Ferroelectric-1}
	S'=\int d^Dr\left[\frac{1}{2}(\nablab\vec{P})^2+\frac{a}{2}\vec{P}^2+\frac{b}{8}(\vec{P}^2)^2-i\nablab\lambda\cdot\vec{P}\right].
\end{equation}
The next step is to approximately integrate out the polarization field and subsequently the Lagrange multiplier field. We do this by performing a shift in one of the directions by a uniform background field $P_0$, assuming the fluctuations around the background field are small enough such that only Gaussian fluctuations are kept. This procedure is equivalent in Feynman diagram language to performing a one-loop calculation of the partition function. Thus, we write $P_1=P_0+\widetilde{P}_1$ where $\widetilde{P}_1$ is a Gaussian fluctuation, treating all $D-1$ remaining transverse fluctuations as Gaussian ones as well. This leads to the approximate Lagrangian, 
\begin{eqnarray}
	\label{Eq:L-Gaussian}
	\mathcal{L}&=&\frac{a}{2}P_0^2+\frac{b}{8}P_0^4+\frac{1}{2}\sum_{j=1}^{D}(\nablab\widetilde{P}_j)^2+\frac{1}{2}\left(a+\frac{3}{2}bP_0^2\right)\widetilde{P}_1^2
	\nonumber\\
	&+&\frac{1}{2}\left(a+\frac{bP_0^2}{2}\right)\sum_{j=2}^{D}\widetilde{P}_j^2-i\sum_{j=1}^{D}\partial_j\lambda\widetilde{P}_j.
\end{eqnarray}
We have omitted terms linear in the fluctuations, since the latter may be absorbed into an external source (like an electric field) and then either set to vanish or be used for a Legendre transformation leading to the same Lagrangian as above. Alternatively, in a symmetry broken regime where $a<0$ the linear terms vanish due to the minimization condition leading to $P_0^2=-2a/b$. 

By integrating out the polarization fluctuations and then the auxiliary field $\lambda$ (see Appendix~\hyperlink{app:RG}{\ref{app:RG}} for calculations), we arrive at the effective action $s_{\rm eff}$ per unit volume, expanded up to quadratic order in $b$:
\begin{eqnarray}\label{Eq:Seff2}
	s_{\rm eff}&\approx&s_0+\frac{a}{2}P_0^2+\frac{b}{8}P_0^4 \nonumber \\
	&+&\frac{N_D}{4}\left(D-\frac{2}{D}+1\right)bP_0^2\int_{0}^{\Lambda}dq\frac{q^{D-1}}{q^2+a}
	\\
	&-&\frac{N_D}{2}\left[\frac{D+7}{8}-\frac{3(D+1)}{2D(D+2)}\right]b^2P_0^4\int_{0}^{\Lambda}dq\frac{q^{D-1}}{(q^2+a)^2},
	\nonumber
\end{eqnarray}
where $s_0$ is an unimportant term not depending on $P_0$ and we have defined $N_D=2^{1-D}\pi^{-D/2}/\Gamma(D/2)$. Note that the integrals in momentum space $\vec{q}$ contain a UV cutoff $\Lambda$. The renormalization group (RG) equations for both $a$ and $b$ are obtained by demanding that $s_{\rm eff}$ is an RG invariant, i.e., $\Lambda d s_{\rm eff}/d\Lambda=0$.
After applying this RG invariance condition for each power of $P_0$ and defining the dimensionless coupling constants, $\hat{a}=\Lambda^{-2}a$ and $\hat{b}=N_D\Lambda^{D-4}b$, we obtain the RG $\beta$-functions, 
\begin{gather}
	\label{Eq:beta-a-b}
	\Lambda\frac{d\hat{a}}{d\Lambda}=-2\hat{a}-\frac{1}{2}\left(D-\frac{2}{D}+1\right)\frac{\hat{b}}{1+\hat{a}}, \\
	\Lambda\frac{d\hat{b}}{d\Lambda}=-(4-D)\hat{b}+\left[\frac{D+7}{2}-\frac{6(D+1)}{D(D+2)}\right]\frac{\hat{b}^2}{(1+\hat{a})^2}. \nonumber
\end{gather}
Note that the obtained RG functions~(\ref{Eq:beta-a-b}) already reveal a significant difference compared to those for unconstrained systems with $N\neq D$ (Appendix~\hyperlink{app:no_constraint}{\ref{app:no_constraint}}). One direct consequence is that for $D=1$ the RG equations become linear, reflecting the fact that there are no transverse directions in this case, which means that no fluctuations exist in the system. This highlights the nontrivial impact of the divergence-free condition on the critical behavior.

For small $\hat{a}$ the theory features an infrared stable fixed point, but the universality class is clearly not the $O(D)$ one. Indeed, linearizing the RG equations~(\ref{Eq:beta-a-b}) near the fixed point leads to the following thermal eigenvalue,
i.e. the RG eigenvalue associated with the relevant direction that tunes the system to criticality, governing the behavior of the correlation length,
\begin{equation}
	\lambda_T=-\frac{8 + D (D^2 + 2D +8)}{D^2 +10D+12},
\end{equation}
which determines the corresponding critical exponent, $\nu=-1/\lambda_T$. Since the number of components of the field $\vec{P}$ is the same as the dimensionality, $D$, of the system, the $\epsilon$-expansion is not suitable for calculating the critical properties. Instead, we have to rely on a fixed dimension approach. For $D=3$, we find $\nu\approx 0.662$.
This is to be compared to the one-loop calculation result for the $O(3)$ universality class in three dimensions, $\nu\approx 0.647$~\cite{Itzykson1991}. Hence, at this order the constraint $\nablab\cdot\vec{P}=0$ causes only a small departure from the usual $O(3)$ universality class for the correlation length scaling.  

\textit{Transverse correlation function --} The zero divergence constraint of the polarization implies that the correlation function $\Delta_{ij}(\vec{r})=\langle P_i(\vec{r})P_j(0)\rangle$ is transverse, $\partial_i \Delta_{ij}(\vec{r})=0$,
or, equivalently, $q_i\tilde{\Delta}_{ij}(\vec{q})=0$ in momentum space. To calculate it, we introduce a quadratic term $\alpha\lambda^2/2$ for the auxiliary field $\lambda$ in Eq.~\eqref{Eq:S-Ferroelectric-1}. Integrating out $\lambda$ generates the term $(\nablab\cdot \vec{P})^2/(2\alpha)$ in the effective action. Hence, by inverting the kernel $\left(q^2+a\right)\delta_{ij}+{q_iq_j}/{\alpha}$ and setting the tuning parameter $\alpha \to 0$ (see Appendix~\hyperlink{app:no_constraint}{\ref{app:no_constraint}}), the free propagator becomes,
\begin{equation}
	\label{Eq:Prop}
	\tilde{\Delta}_{ij}^{(0)}(\vec{q})=\frac{1}{q^2+a}\left(\delta_{ij}-\frac{q_iq_j}{q^2}\right). 
\end{equation} 
The corresponding free propagator in real space, $\Delta _{ij}^{\left( 0\right)}\left( 
\vec{r}\right)$, is given in Appendix~\hyperlink{app:PropReal}{\ref{app:PropReal}}. The calculation of the tadpole correction to the propagator~\eqref{Eq:Prop}, see Appendix~\hyperlink{app:Tadpole}{\ref{app:Tadpole}}, shows that any fluctuation correction to the mass term is transverse. It also recovers the same RG equation for $\hat{a}$ in \eqref{Eq:beta-a-b} as obtained using the effective action approach.

At the next step, we calculate the two-loop correction required to estimate the anomalous dimension $\eta$ of a system under divergence-free conditions.

\textit{Two-loop calculation of $\eta$ --} To derive the value of the anomalous dimension of the polarization at the two-loop order, we represent the interaction term in~(\ref{Eq:S-Ferroelectric}) as
\begin{equation}
\mathcal{L}_{\rm int}=\frac{1}{8}b_{kmnp}P_{k}P_{m}P_{n}P_{p},
\end{equation}
where, for an isotropic system, 
\begin{equation}
b_{kmnp}=\frac{b}{3}\left( \delta _{km}\delta _{np}+\delta _{kn}\delta
_{mp}+\delta _{kp}\delta _{mn}\right). 
\end{equation}

Performing the calculations for the two-point function to order $b^2$ (see Appendix~\hyperlink{app:Sunrise}{\ref{app:Sunrise}}) yields for the two-loop propagator in $D=3$, 
\begin{eqnarray}\label{Eq:prop-q}
\tilde{\Delta}_{ij}\left( \vec{q}\right)  &\equiv&\left( \delta
_{ij}-\frac{q_{i}q_{j}}{q^{2}}\right)\tilde{\Delta}\left( \vec{q}\right), \\
\tilde{\Delta}\left( \vec{q}\right)  &=&\frac{1}{q^{2}+a} \left[ 1- \frac{b^{2}}{ 4\pi^{2}}\frac{C_1\ln\left( c\sqrt{a}\right)+C_2 \frac{q^{2}}{a}+{\it o}(q^2)}{q^{2}+a} \right], \nonumber 
\end{eqnarray}
where $C_1= 11/6$, $C_2\approx 0.04$, and $c$ contains the UV cutoff of the real-space integrals, i.e. $c\to 0$.

Next, we rewrite the propagator~(\ref{Eq:prop-q}) in the form
\begin{eqnarray}\label{Eq:prop-q2}
\tilde{\Delta}\left( \vec{q}\right) \simeq\left( q^{2}+a + \frac{b^{2}}{a}\frac{C_1 a\ln\left( c\sqrt{a}\right)+C_2 q^{2}}{4\pi^{2} } \right)^{-1}. 
\end{eqnarray}

Using this expression, we derive the anomalous dimension $\eta$  following the approach in~\cite{zinn2021quantum} as,
\begin{equation} \label{eq:eta}
\eta=-\Lambda\frac{\partial}{\partial{\Lambda}}{\ln Z}; \qquad Z \left( a,b\right) = 1 + \frac{C_2 }{4\pi^{2} } \frac{b^{2}}{a},
\end{equation}
where $Z$ is the field renormalization constant, and $\eta$ is calculated at the fixed point. 
More precisely, the process of renormalization involves the replacement of bare quantities by renormalized ones, which are overall independent of $\Lambda$ and are given as a perturbative expansion in terms of the bare parameters of the theory. The requirement that all renormalized correlation functions are $\Lambda$-independent yields the RG functions \cite{zinn2021quantum}. In the case of $\eta$, we have explicitly (see Appendix~\hyperlink{app:EtaDerive}{\ref{app:EtaDerive}}),
\begin{equation}\label{Eq:eta2loops}
\frac{\eta}{2}=\left( 1+ C_2 \pi^{2} \frac{\hat{b}_*^{2}}{\hat{a}_*} \right)^{-1}-1.
\end{equation}
Substituting the fixed point values, $\hat{a}_*=-25/77$ and $\hat{b}_*=5/17$, calculated from Eqs.~(\ref{Eq:beta-a-b}) for $D=3$, we obtain $\eta\approx 0.239$. 
This value is significantly larger than the ones typically obtained from the LGW theory, thus highlighting the impact of the constraint $\nablab\cdot\vec{P}=0$.  

The value of the exponent $\eta$ can be experimentally verified in nanoscale ferroelectrics via finite-size scaling of the dielectric susceptibility $\chi$ near the critical temperature, where $\chi\left( \vec{r}\right) \propto \int d^Dr \langle P\left( 0\right) P\left( \vec{r}\right) \rangle = \tilde{\Delta}_{ij}\left( \vec{q}=0\right)$ decays as $\sim r^{2-D-\eta}$. In confined geometries of the linear size $L$, this leads to the finite-size scaling form $\chi\left( L\right)\sim L^{2-\eta}$~\cite{Fisher1972}, allowing extraction of $\eta$ from the size dependence of dielectric response (e.g., on film thickness or nanoparticle diameter) near the Curie point. This approach has been validated in magnetic systems~\cite{Koch2009,PhamPhu2009,Baaziz2015}.

\textit{Conclusions and final remarks --} 
We have shown that imposing a local divergence-free constraint on the polarization field drives a ferroelectric system into a universality class that lies beyond the conventional LGW framework. This profoundly modifies the configuration space, restricting it to loop-like polarization structures and thereby altering the critical behavior. The RG analysis demonstrates that the polarization correlation function features a large anomalous dimension in three dimensions, $\eta=0.239$, which significantly exceeds that of ordinary $O(3)$ LGW models.

This unconventional scaling originates from a naturally induced $U(1)$ gauge symmetry, which becomes explicit when the constraint is resolved as $\vec{P}=\nablab\times\vec{A}$, rendering the effective theory manifestly gauge invariant. 
In this sense, the zero divergence constraint is understood as a local conservation law corresponding to a local gauge symmetry. This leads to the emergence of a gauge field and therefore to a large value of $\eta$. Usually, the argument runs the opposite way, with a gauge symmetry yielding a local conservation law, one example being current conservation in electrodynamics. From a field theoretic perspective, at the critical point current conservation implies that currents {\it do not} have an anomalous dimension \cite{collins1984renormalization}. This fact is a consequence of the scale invariance of the critical point. The opposite is true here: the order parameter, viewed as a conserved quantity, has a large anomalous dimension.  

Looking ahead, ferroelectric systems that naturally enforce local polarization constraints~\cite{Lukyanchuk2025} open a route for exploring gauge-symmetry mediated universality and constraint-induced critical behavior in classical systems.

\section*{Acknowledgements}

S.K. acknowledges the support from the Philipp Schwartz Initiative of the Alexander von Humboldt Foundation. A.S. acknowledges support from the Norwegian Research
Council through Grant No. 262633, “Center of Excellence
on Quantum Spintronics” and Grant No. 323766, as well as
COST Action CA21144 “Superconducting Nanodevices and
Quantum Materials for Coherent Manipulation.”

\section*{Data availability}

No data were created or analyzed in this study.

\section*{Appendices}

\phantomsection\customlabel{app:Helicity}{A}
\hypertarget{app:Helicity}{}
\textit{Appendix~A: Arnold's helicity theorem --}
The helicity, or Hopf invariant, of a divergence-free field $\vec{P}$ defined in a volume $V$ can be expressed through its divergenceless vector potential $\vec{A}$, i.e. $\vec{P}=\nablab\times\vec{A}$, $\nablab\cdot\vec{A}=0$, as~\cite{Arnold2021} 
\begin{equation}
	\label{Eq:helicity}
	\mathscr{H}=\int_V d^3r \,\vec{P}\cdot\vec{A}.
\end{equation}
The $P$-loop representation of polarization vector~(\ref{Eq:P-loop}) assumes the vector potential $\vec{A}(\vec{r})$,
\begin{eqnarray}
	\label{Eq:Apotential}
	\vec{A}&=&-\int_V d^3r' \frac{\nablab'\times\vec{P}(\vec{r'})}{4\pi\lvert \vec{r}-\vec{r'} \rvert} =\int_V d^3r' \frac{\left( \vec{r}-\vec{r'}\right)\times\vec{P}(\vec{r'})}{4\pi\lvert \vec{r}-\vec{r'} \rvert^3} \nonumber \\
    &=&\frac{1}{4\pi}\sum_a  q_a\int_V d^3r' \frac{\left( \vec{r}-\vec{r'}\right)\times\oint_{L_a}  d \gammab^{(a)} \delta^3 \left( \vec{r'} - \gammab^{(a)} \right)}{\lvert \vec{r}-\vec{r'} \rvert^3} \nonumber
    \\
    &=&-\frac{1}{4\pi}\sum_a  q_a \oint_{L_a} \frac{ d \gammab^{(a)} \times \left( \vec{r}-\gammab^{(a)}\right) }{\lvert \vec{r}-\gammab^{(a)}\rvert^3}. 
\end{eqnarray}

Then, the helicity~(\ref{Eq:helicity}) becomes
\begin{eqnarray}
	\label{Eq:helicity-loops}
	\mathscr{H}&=&-\frac{1}{4\pi}\int_V d^3r \,\sum_b  q_b \oint_{L_b}  d \gammab^{(b)} \delta^3 \left( \vec{r} - \gammab^{(b)} \right)\nonumber\\
    &&\cdot\sum_a  q_a \oint_{L_a} \frac{ d \gammab^{(a)} \times \left( \vec{r}-\gammab^{(a)}\right) }{\lvert \vec{r}-\gammab^{(a)}\rvert^3}\\
    &=&\frac{1}{4\pi}\sum_{a,b} q_a  q_b \oint\limits_{L_a} \oint\limits_{L_b}     \frac{ \left[ d \gammab^{(b)} \times d \gammab^{(a)} \right]\cdot \left( \gammab^{(b)}-\gammab^{(a)}\right)}{\lvert \gammab^{(b)}-\gammab^{(a)}\rvert^3}.\nonumber
\end{eqnarray}
Hence, the helicity is expressed in terms of the Gauss linking integral. It represents the linking number of the curves $L_a$ and $L_b$, and corresponds to the average linking of the polarization field streamlines~\cite{Lukyanchuk2025}, $\mathscr{L}_{\vec{P}}=\mathscr{H}$.

\medskip
\phantomsection\customlabel{app:RG}{B}
\hypertarget{app:RG}{}
\textit{Appendix~B: Renormalization group calculations --}
Integrating out the Gaussian fluctuations of the polarization $\widetilde{P}_j$ in~(\ref{Eq:L-Gaussian}) yields the following effective action,
\begin{eqnarray}
	\label{Eq:Seff}
	S_{\rm eff}&=&V\left(\frac{a}{2}P_0^2+\frac{b}{8}P_0^4\right)+\frac{1}{2}{\rm Tr}\ln\left(-\nabla^2+a+\frac{3}{2}bP_0^2\right)
	\nonumber\\
	&+&\frac{D-1}{2}{\rm Tr}\ln\left(-\nabla^2+a+\frac{bP_0^2}{2}\right)\nonumber\\
	&+&\frac{1}{2}\int d^Dr\int d^Dr'\left[\vphantom{\sum_{j=2}^{D}}\partial_1\lambda(\vec{r})G_L(\vec{r}-\vec{r}')\partial_1'\lambda(\vec{r}')
	\right.\nonumber\\
	&+&\left.\sum_{j=2}^{D}\partial_j\lambda(\vec{r})G_T(\vec{r}-\vec{r}')\partial_j'\lambda(\vec{r}')\right],
\end{eqnarray}
where $V$ is the (infinite) volume, and $G_L(\vec{r}-\vec{r}')$ and $G_T(\vec{r}-\vec{r}')$ have Fourier transforms, 
\begin{equation}
	G_L(\vec{q})=\frac{1}{q^2+a+\frac{3bP_0^2}{2}},\quad G_T(\vec{q})=\frac{1}{q^2+a+\frac{bP_0^2}{2}}. 
\end{equation} 
Next, we integrate out the Lagrangian multiplier field, which gives the following effective action $s_{\rm eff}$ per unit volume (which in this case corresponds to the free energy density) written in terms of momentum space variables, 
\begin{eqnarray}
\label{Eq:Seff-1}
	s_{\rm eff}&=&\frac{a}{2}P_0^2
    +\frac{b}{8}P_0^4+\left(\frac{D}{2}-1\right)\int_q\ln\left(q^2+a+\frac{bP_0^2}{2}\right) \nonumber \\
	&+&\frac{1}{2}\int_q\ln\left[q^2\left(q^2+a+\frac{bP_0^2}{2}\right)
	+bP_0^2\sum_{j=2}^{D}q_j^2\right],
\end{eqnarray}
where $\int_q\equiv\int\frac{d^Dq}{(2\pi)^D}$. The above form of the effective action reveals a different pattern of would-be Goldstone modes. Typically, without the constraint $\nablab\cdot \vec{P}=0$, the minimum of the effective potential in the spontaneous symmetry breaking regime would yield $D-1$ gapless transverse modes and one longitudinal gapped mode, as required by the $O(D)$ symmetry. At one loop, the minimum is given by the tree level contribution, $P_0^2=-2a/b$, requiring $a<0$. From Eq. (\ref{Eq:Seff-1}), all modes are gapless as $q\to 0$, a direct consequence of the constraint, which imposes transversality of $\vec{P}$. A gapped mode and the usual Goldstone counting appear only if $q_1\to 0$ faster than the other $\vec{q}$ components, reminiscent of early studies of the critical behavior in systems with dipolar interaction~\cite{Aharony_PhysRevB.8.3363}. 

In order to obtain the one-loop RG equations, we expand $s_{\rm eff}$ up to the second order in $b$, which results in the expression~(\ref{Eq:Seff2}) for the effective action.

\medskip
\phantomsection\customlabel{app:no_constraint}{C}
\hypertarget{app:no_constraint}{}
\textit{Appendix~C: RG functions with no constraint --} In the absence of the $\nablab\cdot \vec{P}=0$ constraint, the one-loop RG equations~(\ref{Eq:beta-a-b}) would acquire the usual $O(N)$ form,
\begin{gather}
	\label{Eq:beta-a-b-noconstraint}
	\Lambda\frac{d\hat{a}}{d\Lambda}=-2\hat{a}-\frac{N+2}{2}\frac{\hat{b}}{1+\hat{a}}, \\
	\Lambda\frac{d\hat{b}}{d\Lambda}=-(4-D)\hat{b}+\frac{N+8}{2}\frac{\hat{b}^2}{(1+\hat{a})^2}. \nonumber
\end{gather}
It is clearly seen that imposing $N = D$ directly does not lead to Eqs.~(\ref{Eq:beta-a-b}), which highlights the nontrivial impact of the constraint on the critical properties of the system. 

To illustrate the effect of the constraint, one may keep the tuning parameter $\alpha$ in the free propagator~\eqref{Eq:Prop} as:
\begin{equation}
	\tilde{\Delta}_{ij}^{(0)}(\vec{q}\,;\alpha)=\frac{1}{q^2+a}\left(\delta_{ij}-\frac{q_iq_j}{(1+\alpha) q^2+\alpha a}\right), 
\end{equation} 
thus setting $\alpha \to 0$ results in the considered constrained case, while $\alpha \to \infty$ corresponds to the absence of the constraint. The $\alpha$-dependent RG $\beta$-functions read
\begin{eqnarray}
	\Lambda\frac{d\hat{a}}{d\Lambda}&=&-2\hat{a}-\frac{1}{2}\left(D-\frac{2}{D}+1\right)\frac{\hat{b}}{1+\hat{a}} - \frac{D+2}{2D}\frac{\frac{\alpha}{1+\alpha}\hat{b}}{1+\frac{\alpha}{1+\alpha}\hat{a}}, \nonumber \\
	\Lambda\frac{d\hat{b}}{d\Lambda}&=&-(4-D)\hat{b}+\left[\frac{D+7}{2}-\frac{6(D+1)}{D(D+2)}\right]\frac{\hat{b}^2}{(1+\hat{a})^2} \nonumber \\ 
    &+& \frac{\alpha}{1+\alpha}\frac{4(D-1)}{D(D+2)}\frac{\hat{b}^2}{(1+\hat{a})(1+\frac{\alpha}{1+\alpha}\hat{a})} \\
    &+&\frac{1}{2}\frac{\alpha^2}{(1+\alpha)^2}\left[1+\frac{4}{D}+\frac{12}{D(D+2)}\right]\frac{\hat{b}^2}{(1+\frac{\alpha}{1+\alpha}\hat{a})^2}, \nonumber
\end{eqnarray}
and reduce to Eqs.~(\ref{Eq:beta-a-b}) and~(\ref{Eq:beta-a-b-noconstraint}) with $N = D$ in two limit cases $\alpha \to 0$ and $\alpha \to \infty$, respectively.

\medskip
\phantomsection\customlabel{app:PropReal}{D}
\hypertarget{app:PropReal}{}
\textit{Appendix~D: Free polarization proparator in real space --}
The constraint $\nablab\cdot\vec{P}=0$ imposes that in the absence of interactions the polarization propagator is given in momentum space by Eq.~(\ref{Eq:Prop}). To obtain the corresponding propagator in real space, we rewrite the propagator as, 
\begin{equation}
	\label{Eq:Prop-1}
	\tilde{\Delta}_{ij}(\vec{q})=\frac{\delta_{ij}}{q^2+a}+\frac{q_iq_j}{a}\left(\frac{1}{q^2+a}-\frac{1}{q^2}\right), 
\end{equation}
and introduce the function,
\begin{eqnarray}
	C(r;a)&=&\int\frac{d^Dq}{(2\pi)^D}\frac{e^{i\vec{q}\cdot\vec{r}}}{q^2+a}
	\nonumber\\
	&=&\frac{1}{(2\pi)^{D/2}}\left(\frac{\sqrt{a}}{r}\right)^{\frac{D-2}{2}}K_{\frac{D-2}{2}}(\sqrt{a}r),
\end{eqnarray}
where $K_\nu(x)$ is the modified Bessel function of the second kind. At large distances,
\begin{equation}
   	C(r;a)\approx\frac{1}{2\sqrt{a}}\left(\frac{\sqrt{a}}{2\pi r}\right)^{(D-1)/2} e^{-\sqrt{a}r}, 
\end{equation}
which for $D=3$ holds exactly. Then, in real space,
\begin{eqnarray}\label{Eq:PropD}
&&\Delta_{ij}^{(0)}\left( \vec{r}\right) =
C(r;a)\delta_{ij}-\frac{1}{a}\partial_i \partial_j \left[ C(r;a)-C(r;0) \right]
\nonumber \\
&=&\frac{1}{4\pi ar^{D}} \left( \frac{1}{2\pi} \right)^{\frac{D-3}{2}} \nonumber \\ &\times& \left\{ \delta _{ij}\left[
e^{-\sqrt{a}r}(\sqrt{a}r)^{\frac{D-3}{2}}\left( (\sqrt{a}r)^2+\sqrt{a}r+\frac{D-1}{2}\right) \right.\right. \nonumber \\
&-&\left.\left.\frac{2^{\frac{D}{2}}\Gamma(\frac{D}{2})}{\sqrt{2\pi}}\right] +\frac{x_{i}x_{j}}{%
r^{2}}\left[D\frac{2^{\frac{D}{2}}\Gamma(\frac{D}{2})}{\sqrt{2\pi}}-
e^{-\sqrt{a}r}(\sqrt{a}r)^{\frac{D-3}{2}}\right.\right. \nonumber \\
&\times&\left.\left.\left( (\sqrt{a}r)^2+D\sqrt{a}r+\frac{(D-1)(D+3)}{4}\right) \right] \right\},
\end{eqnarray}

For $D=3$, this expression simplifies to
\begin{eqnarray}\label{Eq:PropD3}
\Delta_{ij}^{(0)}\left( \vec{r}\right) &=&\frac{1}{4\pi ar^{3}}\left\{ \left[
e^{-\sqrt{a}r}\left( ar^{2}+\sqrt{a}r+1\right) 
-1\right] \delta _{ij}\right. \nonumber \\ &-&\left.\left[
e^{-\sqrt{a}r}\left( ar^{2}+3\sqrt{a}r+3\right) -3\right] \frac{x_{i}x_{j}}{%
r^{2}}\right\},
\end{eqnarray}
allowing us to calculate the two-loop propagator~(\ref{Eq:prop-q}).

\medskip
\phantomsection\customlabel{app:Tadpole}{E}
\hypertarget{app:Tadpole}{}
\textit{Appendix~E: Tadpole correction to the propagator --}
The one-loop propagator is given by, 
\begin{eqnarray}
    \label{Eq:G-one-loop}
    \widetilde{\Delta}_{in}(\vec{q})&=&\widetilde{\Delta}^{(0)}_{in}(\vec{q})-\frac{3}{2}\widetilde{\Delta}_{ij}^{(0)}(\vec{q})\widetilde{\Delta}_{mn}^{(0)}(\vec{q})b_{jklm}\int_k\widetilde{\Delta}_{kl}^{(0)}(\vec{k})
    \nonumber\\
    &=&\left(\delta_{in}-\frac{q_iq_n}{q^2}\right)\left[\frac{1}{q^2+a}\right.\nonumber\\
    &-&\left.\frac{b}{2}\left(D-\frac{2}{D}+1\right)\frac{1}{(q^2+a)^2}
\int_k\frac{1}{k^2+a}\right]
    \nonumber\\
    &=&\frac{1}{q^2+a}\left(\delta_{in}-\frac{q_iq_n}{q^2}\right)\nonumber\\
    &\times&\left[\vphantom{\frac{1}{q^2+a}}~1
    -\frac{b}{2(q^2+a)}\left(D-\frac{2}{D}+1\right)\int_k\frac{1}{k^2+a}\right]
    \nonumber\\
    &=&\frac{1}{\Gamma^{(2)}(\vec{q})}\left(\delta_{in}-\frac{q_iq_n}{q^2}\right),
\end{eqnarray}

where $\int_k\equiv\int\frac{d^Dk}{(2\pi)^D}$ and
\begin{equation}
    \Gamma^{(2)}(\vec{q})=q^2+a+\frac{b}{2}\left(D-\frac{2}{D}+1\right)\int_k\frac{1}{k^2+a}.
\end{equation}
The above result shows explicitly what we should expect, that any fluctuation correction to the mass term should be transverse. No renormalizations in the longitudinal direction takes place. We can thus define the renormalized quantity, $a_R=\Gamma^{(2)}(0)$ and obtain the $\vec{P}^2$ operator insertion leading to the critical exponent $\nu$ as $\partial a_R/\partial a$. Introducing the dimensionless bare coupling $\hat{a}=\Lambda^{-2}a$ and demanding that $a_R$ is independent of the cutoff $\Lambda$ recovers the RG equation for $\hat{a}$ in \eqref{Eq:beta-a-b}, which was obtained using the effective action approach.

\medskip
\phantomsection\customlabel{app:Sunrise}{F}
\hypertarget{app:Sunrise}{}
\textit{Appendix~F: Two-loop propagator --} The two-point function to the order $b^2$ in momentum space, 
\begin{equation}\label{Eq:2point}
\tilde{\Delta}_{ij}\left( \vec{q}\right) =\tilde{\Delta}_{ij}^{\left(
0\right) }\left( \vec{q}\right) +\frac{3}{2}\tilde{\Delta}_{ik}^{\left(
0\right) } \tilde{\Delta}_{jl}^{\left( 0\right)
} \mathcal{A}_{kl}\left( \vec{q}\right) ,
\end{equation}
\begin{center}
\begin{fmffile}{sunrise}
\begin{fmfgraph*}(120,60)
  \fmfleft{i}
  \fmfright{o}

  \fmf{plain}{i,v1}
  \fmf{plain}{v2,o}

  \fmf{plain,left=1,tension=0.2}{v1,v2}
  \fmf{plain,left=0,tension=0.2}{v1,v2}
  \fmf{plain,right=1,tension=0.2}{v1,v2}

  \fmfdot{i,o,v1,v2}

  \fmfv{label=$i\qquad k \quad n \qquad\quad t \quad l \qquad j$, label.angle=60, label.dist=2pt}{i}
  \fmfv{label=$\ m\quad\ s$, label.angle=60, label.dist=20pt}{v1}
  \fmfv{label=$\ p\qquad\ u$, label.angle=-130, label.dist=15pt}{v2}
\end{fmfgraph*}
\end{fmffile}
\end{center}
includes the contribution from the sunrise diagram above,
\begin{equation}\label{Eq:Sunrise_diag}
\mathcal{A}_{kl}\left( \vec{q}\right) =b_{kmnp}b_{lstu}\int \Delta _{ms}^{\left( 0\right) } \Delta _{nt}^{\left( 0\right) }
\Delta _{pu}^{\left( 0\right) } e^{-i\vec{q}\cdot\vec{r}}d^{D}r.
\end{equation}
Here, $\Delta _{ij}^{\left( 0\right)}\left( 
\vec{r}\right)$ is the free polarization propagator in real space, see Appendix~\hyperlink{app:PropReal}{\ref{app:PropReal}}. Note that in the massless theory ($a=0$) the diagram~(\ref{Eq:Sunrise_diag}) is logarithmically divergent for $D=3$.
This issue does not arise in massive theory. 

Although the diagram $\mathcal{A}_{kl}$~(\ref{Eq:Sunrise_diag}) representing the amputated correlation function is not transverse, the full second-order propagator~(\ref{Eq:2point}) is transverse as required by the constraint. Performing the integral in~(\ref{Eq:Sunrise_diag}) and substituting into~(\ref{Eq:2point}) yields for $D=3$ the expression~(\ref{Eq:prop-q}) with $C_1= 11/6$, $C_2= \left( 792 \ln2-315\ln3-181 \right)/540$.

\medskip
\phantomsection\customlabel{app:EtaDerive}{G}
\hypertarget{app:EtaDerive}{}
\textit{Appendix~G: Explicit derivation of the anomalous dimension --}
The first nonvanishing contribution to the field renormalization constant, $Z$, is given by the coefficient of the $q^2$ term in the renormalized two-point vertex function corresponding to the two-loop propagator~(\ref{Eq:prop-q},\ref{Eq:prop-q2}) as
\begin{equation}
\quad Z \left( a,b\right) =-\frac{1}{\tilde{\Delta}^2}\frac{\partial{\tilde{\Delta}}}{\partial{q^2}}\bigg|_{q^2=0}= 1 + \frac{C_2 }{4\pi^{2} } \frac{b^{2}}{a}.
\end{equation}

The RG function $\eta\left( a,b\right)$ is given by~\cite{zinn2021quantum}, 
\begin{equation} \label{eq:RG-eta}
\eta \left( a,b\right)=-\Lambda\frac{\partial}{\partial{\Lambda}}\bigg|_{b_R {\text{ fixed}}}{\ln Z}\simeq-\frac{1}{Z}\frac{C_2 }{4\pi^{2} }\frac{2 b^2 }{a }.
\end{equation}
Note that the differentiation with respect to $\Lambda$ is performed keeping the (renormalized) coupling constant, but not the mass, fixed. Specifically, we replace the bare quantities $a,b$ by $\Lambda$-independent renormalized ones $a_R,b_R$, which are then given as a perturbative expansion in terms of the bare parameters of the theory. Next, recalling the dimensionless coupling constants, $\hat{a}=\Lambda^{-2}a$ and $\hat{b}=N_D\Lambda^{D-4}b$, gives $b^2/a=N_D^{-2}\Lambda^{2(3-D)}\hat{b}^2/\hat{a}$, or, in three dimensions, $b^2/a=N_3^{-2}\hat{b}^2/\hat{a}=4\pi^4\hat{b}^2/\hat{a}$.
The values of the dimensionless coupling constants at the fixed point, $\hat{a}_*$ and $\hat{b}_*$, are obtained from the RG equations~(\ref{Eq:beta-a-b}). 
We solve $\Lambda d\hat{b}/d\Lambda=0$ to get $\hat{b}_*$ and then tune the mass parameter $\hat{a}$ to criticality using the RG $\beta$-function for $\hat{a}$. This yields
\begin{equation}
\hat{a}_*=\frac{\left( D-4\right) \left( D+2\right) ^{2}}{D^{3}+2\left( D+2\right)^{2}};\,\,
\hat{b}_*=\frac{2D\left( 4-D\right) \left( D+2\right) }{\left( D-1\right) \left(
D^{2}+10D+12\right) },
\end{equation}
resulting in $\hat{a}_*=-25/77$ and $\hat{b}_*=5/17$ for $D=3$. Substituting the fixed point values to the RG function~(\ref{eq:RG-eta}) leads to the expression~(\ref{eq:eta}), and, eventually, to the numerical value $\eta\approx 0.239$.

%

\onecolumngrid

\clearpage
\twocolumngrid

\end{document}